\documentclass[12pt,preprint]{aastex}

\def\kms{$\rm km\, s^{-1}$}
\def\cm3{$\rm cm^{-3}$}
\def\Ts{$\rm T_{*}$}
\def\Vs{$\rm V_{s}$}
\def\n0{$\rm n_{0}$}
\def\B0{$\rm B_{0}$}
\def\ne{$\rm n_{e}$}
\def\Te{$\rm T_{e}$}
\def\Tgr{$\rm T_{gr}$} 
\def\Tgas{$\rm T_{gas}$}

\def\mum{$\mu$m}
\def\agr{a$_{gr}$}

\begin{document}

\shorttitle{Silicates in D-type symbiotic stars}
\shortauthors{Angeloni et al.}

\title{Silicates in D-type symbiotic stars:\\an ISO overview}

\author{R. Angeloni\altaffilmark{1}, S. Ciroi, and P. Rafanelli}
\affil{Dipartimento di Astronomia, University of Padova, \\Vicolo dell'Osservatorio 2,
    I-35122 Padova, Italy}
\email{angeloni@pd.astro.it, ciroi@pd.astro.it, piraf@pd.astro.it}
\and

\author{M. Contini\altaffilmark{2}}
\affil{School of Physics and Astronomy, Tel-Aviv University, Tel-Aviv 69978 Israel}
\email{contini@post.tau.ac.il}
\altaffiltext{1}{School of Physics and Astronomy, Tel-Aviv University.}
\altaffiltext{2}{Dipartimento di Astronomia, University of Padova.}
\begin{abstract}
We investigate the IR spectral features of a sample of D-type symbiotic stars in order to constrain the emitting properties of coupled dust-gas particles across the whole system. In particular, analyzing unexploited ISO-SWS data, deriving the basic observational parameters of dust bands, and comparing them with respect to those observed in other astronomical sources, we try to highlight the effect of environment on grain chemistry and physics. We find strong amorphous silicate emission bands at $\sim$10 \mum\ and $\sim$18 \mum\ in a large fraction of the sample. The analysis of the $\sim$10 \mum\ band, along with a direct comparison with several astronomical sources, reveals that silicate dust in symbiotic stars shows features between the characteristic circumstellar environments and the interstellar medium. This indicates an increasing reprocessing of grains in relation to specific symbiotic behavior of the objects. A correlation between the central wavelength of the $\sim$10 and $\sim$18 \mum \, dust bands is found. By the modeling of IR spectral lines we investigate also dust grains conditions within the shocked nebul\ae{}. Both the unusual depletion values and the high sputtering efficiency might be explained by the formation of SiO molecul\ae{}, which are known to be a very reliable shock tracer. We conclude that the signature of dust chemical disturbance due to symbiotic activity should be looked for in the outer, circumbinary, expanding shells where the environmental conditions for grain processing might be achieved. Symbiotic stars are thus attractive targets for new mid-infrared and $mm$ observations.
\end{abstract}

\section{Introduction}
Symbiotic systems (SS) are currently understood as interacting binaries composed of
a compact star, generally but not necessarily a white dwarf (WD), a cool giant star, which is at the origin of dust formation and ejection episodes, and different emitting gas and dust nebul\ae{}. In the past years, theoretical models (Girard \& Willson 1987, Kenny \& Taylor 2005)
as well as observations (Nussbaumer et al. 1995) have definitively shown that both the hot and cool stars lose mass through strong stellar winds
which collide within and outside the system, hence creating a complex network of
wakes and shock fronts which result in a complicated structure of gas 
and dust nebul\ae{} (Nussbaumer 2000).

Two main phases characterize the symbiotic activity: the \textit{quiescent phases},
during which the hot component releases its energy approximately at a constant
rate and spectral distribution, and the \textit{active phases}, characterized by
a significant change of the hot component radiation. 
Near infrared (NIR) photometry revealed the continuing presence of a cool variable star in these systems (Wallerstein et al. 1984 and references therein) both before and after an active phase, hence indicating that the eruption episodes do not significantly affect the cool component. Contemporary UV and IR observations showed that while the UV flux decreases during the quiescent phases, the IR remains virtually constant, again suggesting that the IR source is not directly involved in symbiotic activity (Friedjung et al. 1984). \\
Nevertheless recent works, both theoretical and observational, highlit that cool stars in SS might behave quite differently with respect to the isolated ones. For instance, it has been suggested that both the red giants and the Mira variables in SS have systematically higher mass-loss rates than typical galactic giants and Mir\ae{} (Mikolajewska 2000; Mikolajewska et al. 2002), and that the cool giants in SS are likely to rotate much faster than the isolated ones, leading to an enhanced magnetic activity as a main result, and probably to inhomogeneous mass-loss and asymmetric dust-ejection episodes (Soker 2002).

In the '70s, on the basis of near-infrared (NIR) colours, SS were classified in S and D types (Webster \& Allen 1975) according to whether the cool star (S-type) or dust (D-type) dominates the 1-4 \mum \, spectral range.
A peculiar aspect of the symbiotic phenomenon in D types pertains therefore quite naturally to the dusty environment. 
These systems show a broad IR excess which, since the first IR surveys,
has been attributed to emission from circumstellar dust.
It was found that the dust excesses have colour temperatures near 1000 K
(Feast et al. 1983, Whitelock 1987); nevertheless it has been difficult to explain the observed
IR spectrum by a single temperature, and nowadays theoretical models and observations too have been confirming that several dust temperatures should be combined in order to reproduce the NIR-MIR data (Anandarao et al. 1988, Schild et al. 2001, Angeloni et al. 2006a, hereafter Paper I, Angeloni et al. 2006b, hereafter Paper II).

In previous investigations of the SED from SS
(Paper I) we have analyzed the different contributions to the continuum, with special 
emphasis on emission by dust in the IR spectral range.
The results revealed the presence of two dust shells in many systems, pointing out that
multiple dust shells are common in D-type systems:
the inner and hotter ($\sim$ 1000 K) shells appear circumstellar, with a typical
size of 10$^{14}$ cm, while the outer and cooler (T=300 K) with sizes of $\sim$
10$^{15}$ cm are likely to be circumbinary.
Very recent observations by Biller et al. (2006) confirm that in some systems,
such as CH Cyg and HM Sge, even the inner dust shell surrounds both the stars,
interacting in symbiosis with the shocked nebulae. \\

Since the last '60s smooth, broad bands at 10 \mum \, and 18 \mum \, have been observed in the spectra of some late-type, variable stars and then in a wider variety of galactic environments. They were soon associated with silicate dust grains and were assigned to the Si-O stretching and bending modes respectively. Furthermore, although the silicates with various composition all display these spectral features due to Si-O vibrational modes, there are differences in the spectral appearance that allow to constrain the exact composition of the silicate grains, as well. After the ISO "crystalline revolution" (J\"ager et al. 1998) the whole dust profiles have been interpreted in terms of mixed crystalline and amorphous structures (e.g. Bowey \& Adamson 2002, Kemper et al. 2005, Tamanai et al. 2006).

Various attempts have been carrying on to classify the different O-rich dust features on the basis of the spectral profile, by both space-based facilities, like IRAS and ISO (e.g. Little-Marenin \& Little 1990; Sloan \& Price 1995; Posch et al. 1999) and ground-based ones, like UKIRT (Monnier et al. 1998; Speck et al. 2000, hereafter S00). Some studies highlighted also a clear dependence on the environment conditions: for instance, the central wavelength of the peak $\lambda_c$, of the $\sim$10 \mum \, band is known to decrease  from $>$9.8 \mum \,in the circumstellar and diffuse-medium environments to 9.6 \mum \, in the Taurus Molecular Clouds (TMC), while the FWHM increases from ~2.5 \mum\ to 3.3 \mum \,(P\'egouri\'e \& Papoular 1985). These changes in the profiles seem represent the effect of mixing grains from different sources and of chemical and physical processing on the older TMC and Young Stellar Object (YSO) silicates (Bowey \& Adamson 2002, Table 2) and there have been even some works which attempted, on the basis of a long series of laboratory experiments, to relate the ratio of the 10\mum\ and 18 \mum\ band strength with the relative age of silicate (Nuth \& Hecht 1990). \\
An accurate analysis of observational infrared spectra, discriminating the role played by the various environment parameters, is not trivial since not only mineralogical properties but also size, shape and agglomeration of the grains can have an influence on the shape and intensity of the distinctive spectral features. Thus further laboratory studies on the silicate dust bands are essential for an appropriate interpretation of the observed spectra.

In the present work we analyze unexploited ISO data in order to compare the silicate dust features in D-type systems with those of ''normal'', isolated late-type stars and other galactic environments.
We refer to the classification scheme proposed by S00 who, thanks to an infrared survey of more than 140 stars, identified similarities and differences between the features of various types of stars. As we are interested in Mira variables, that correspond to the cool component in D-type SS, which represent $\sim$ 20\% of the SS in the Belczy{\'n}ski et al. (2000) atlas, we will focus on the three dust feature groups proposed by S00 for AGB stars: 1) \textit{broad AGB}, where the feature extends from 8 to about 12.5 \mum \, with little structure; 2) \textit{broad+sil AGB}, which consists of a broad feature with an emerging 9.7 \mum \, silicate bump; and 3) \textit{silicate AGB}, which is the ''classic'' 9.7 \mum \, silicate feature. The \textit{silicate AGB} group is split further into four subgroups, A, B, C, D, which show a sequence of slightly different shapes from broader to narrow silicate features (S00, Fig.1).

Our aim is to study the environmental conditions in SS, both in the dust shells and in the shocked, ionized nebul\ae{}, in order to investigate whether the "symbiosis" might affect the dust behavior, leaving its marks on the spectral profile by changing the chemical and physical properties of the grains, as well as the shell morphology and location, as indicated by Paper I. The sample and data reduction are described in \S 2, while in \S3 the results are presented and discussed. Concluding remarks follow in \S4.
\section{The sample: observations and data reduction}
The sample has been selected by cross-checking Belczy{\' n}ski et al. (2000) atlas with the ISO Archive, and then by including those D-types which showed the best quality observations. It consists of the same 9 SS analyzed in Paper I: 8 D type, namely, R Aqr (RAq), V627 Cas (V6), HM Sge (HM), V1016 Cyg (V1), RR Tel (RR), V835 Cen (V8), H1-36 (H1), BI Bru (BI),  and 1 S-type, CH Cyg (CH), whose actual classification is however still controversial (see Paper I and references therein).

We present the spectra taken with the Short Wavelength Spectrograph (hereafter SWS - de Graauw et al. 1996), used in full-grating scan mode (AOT 01) and covering the wavelength
range between 2.38 and 45.2 \mum.
The spectra come from the ''Uniform database of SWS 2.4-45.4 micron
spectra'' within the Highly Processed Data Products (HPDP) section of the ISO Archive. The database presents a complete set of all valid SWS full-scan 2.4-45.4 \mum \, spectra processed from the last stage of the pipeline software and renormalized as uniformly as possible, thus representing the most processed form available from the ISO archive (see Sloan et al. 2003 for details about the algorithm used to generate the database). However some instrumental artifacts are still present, and in specific ranges and cases the S/N level is not sufficient to derive reliable quantitative measurements: for instance the SWS band 3E at 27.5 - 29 \mum \, is known for its mediocre performance and beyond that wavelengths the spectra quality gradually deteriorates (see Paper II). Anyway this does not affect the analisys of dust amorphous band at all, but it is reflected only in a inability to determine the very accurate values of the atomic line observed parameters. Considering the aim of the present work, we thus preferred to give only upper limits to the emission lines ratios, exploiting HM Sge as a test-case in the analisys of the emitting nebul\ae{} (\S 3.2.1). In the following of the present work we focus on the 7-27 \mum \,spectral range.

In the analysis of the dust bands at $\sim$10 \mum \, and $\sim$18 \mum \, we decided not to subtract any sort of ''continuum'', but to rely on the baseline defined by the task SPLOT within the NOAO-ONEDSPEC IRAF package software (see below) during the fitting procedure. Indeed we verified that, even when the continuum appeared moderately steep, the profile shapes did not change to the extent as to modify the basic band parameters (like $\lambda_{c}$ and FWHM, the full width half maximum) and to prevent the classification of dust features. This has been confirmed \textit{a posteriori} by the classification of the only star that the present sample share with the S00 one, i.e. R Aqr. In fact, in agreement whit S00, we classified it in the \textit{silicate AGB C} group (see $\S$3.1).\\
Furthermore, the identification of the continuum to be subtracted is not trivial: S00, for instance, subtracted a 8 \mum \, normalized 3000K blackbody representing a stellar photosphere, by explicitly assuming that in the 8-13 \mum \, range the continuum is dominated by the stellar contribution. In fact, the modeling of ISO-SWS spectra (Paper I) has shown that in the abovementioned range the continuum emission is predominantly caused by reradiation of hot ($\sim$1000K) dust from the inner shells (r$\sim$10$^{14}$cm). In the longer wavelength domain, i.e. around $\sim$ 20 \mum, the continuum becomes even more composite, being the sum of contributions from at least two dust shells at different temperatures. Any kind of continuum, without a model taking into account the real physical scenario, appears hence artificial and almost meaningless. As our profiles did not change their shape substantially, we chose a direct approach to the dust band analysis.

To measure the dust profile parameters we used two different approaches. 
The dust profiles are clearly asymmetric therefore, by
definition, the central wavelength of the band will not coincide with
the peak position: a gaussian fit is then not able in reproducing the exact,
whole feature profile. 
Nevertheless, as we focused in the central wavelength of the peak
we performed, by means of the NOAO-ONEDSPEC task within the IRAF package
software, a gaussian fit of merely the upper portion of the band,
fairly approximable by gaussian curve. Then, to check this kind of approach, we tried to follow the same method of Bowey \&
Adamson (2002), i.e. to estimate the central wavelength of the
peak and the FWHM of the bands "by eye". A good agreement between the
two methods was found, even if this translates in moderately larger error bars.
Nonetheless it is very unlikely that some not accounted for
physical effects could move the 
real central wavelength of the band peak as much as to shift it out of the error bars. 
It is clear that these uncertainties do not invalidate the
results presented in the next section.
As regards the atomic emission lines, we performed a gaussian fit by means of the above mentioned NOAO-ONEDSPEC task in order to measure their main observational parameters, namely the central wavelength and the FWHM.
\section{Results and discussion}
\begin{figure}[!hp]
\centering
\includegraphics[width=0.45\textwidth]{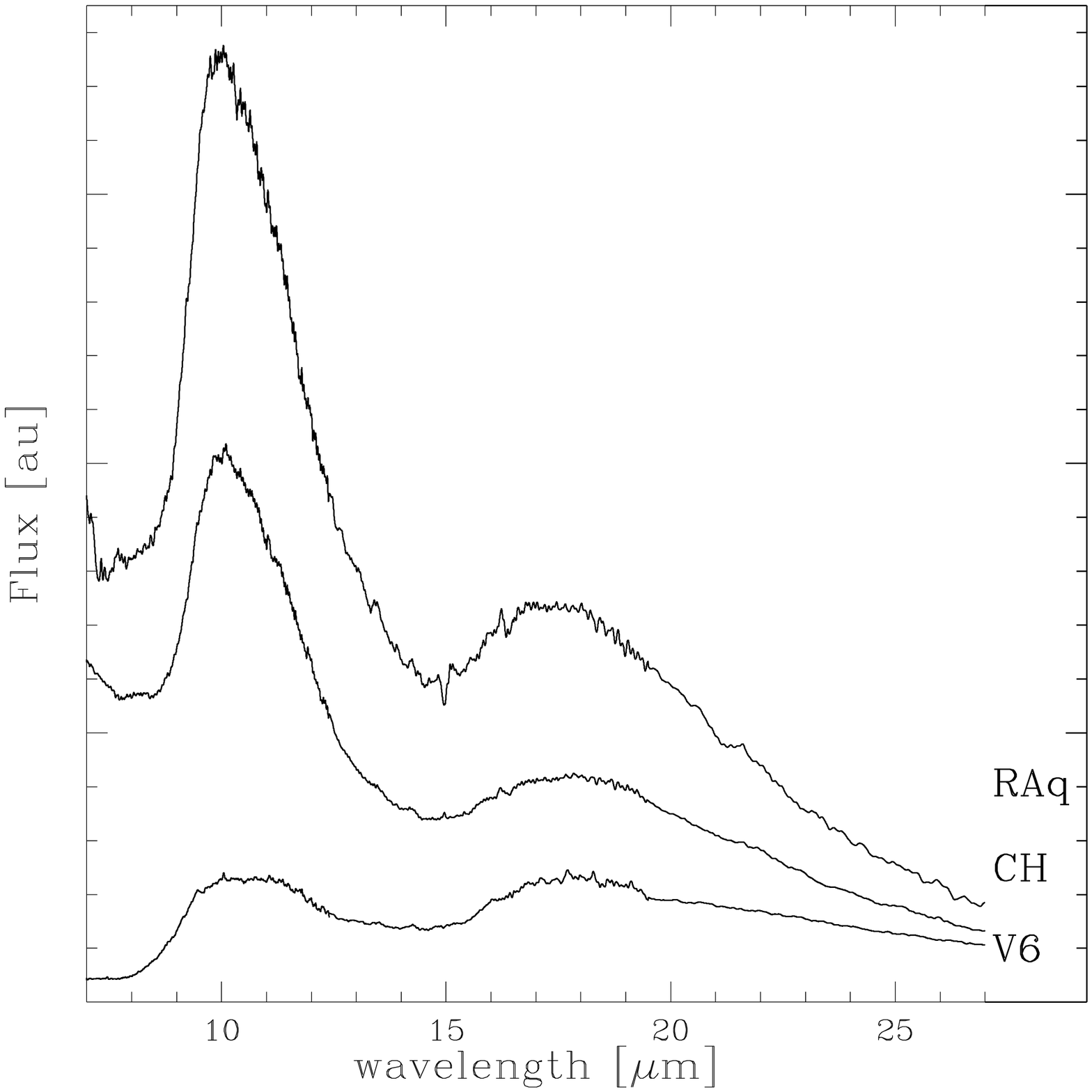}
\includegraphics[width=0.45\textwidth]{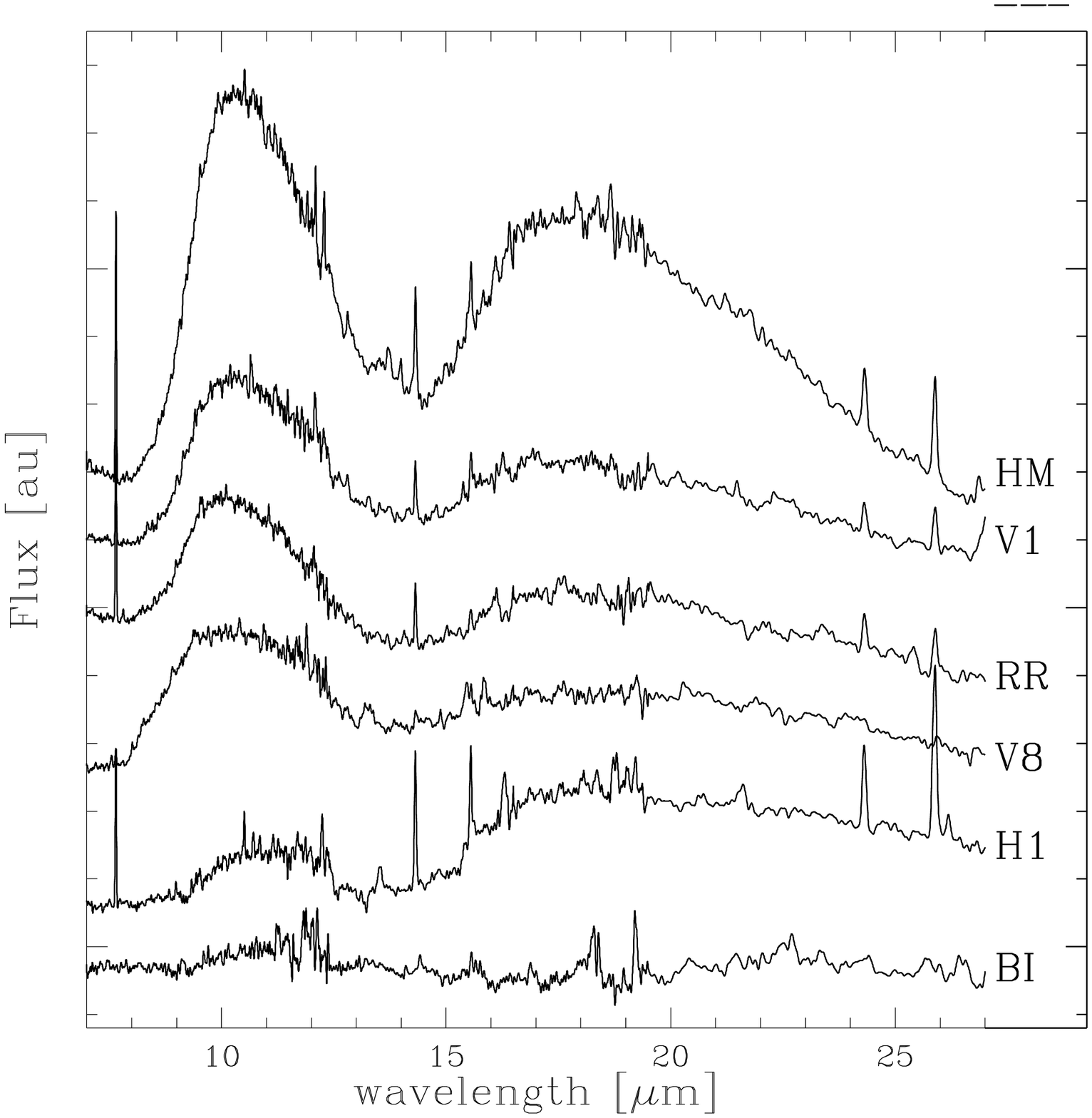}
\caption{The 7-27 \mum \, region of the ISO-SWS spectra for the SS of the sample. Notice the similarities and the differences in the continuum SED, in the dust band profiles at $\sim10$ and $\sim18$ \mum \, (e.g. likely absent in BI Cru, even if the poor S/N of the spectrum doesn't allow to conclusively state about it) and in the intensities of atomic emission lines. \label{fig:sil}}
\end{figure}

In Fig. \ref{fig:sil} we present the 7-22 \mum \, spectra for the SS of our sample. 
The symbiotic character of such systems can be inferred by noticing that on top of the composite stellar-dust continuum, and even overlapped on the dust broad bands, many intense forbidden emission lines are visible. \\
In \S 3.1 we analyze and compare the amorphous silicate spectral profiles which dominate the wavelength range, and in \S 3.2, by exploiting HM Sge as a test-case, we model the IR line spectra in order to constrain the emitting properties of coupled dust-gas particles. In particular, we would like to investigate whether dust grains can survive within the shocked nebul\ae{} and how their chemical properties might be affected by the symbiotic environment as a whole.
\subsection{Silicate spectral features}
The main characteristics of the spectra presented in Fig. \ref{fig:sil} are the two broad, easily recognizable dust bands at 10 \mum \, and 18 \mum, arising from the vibrational modes of Si-O bound. These bands dominate the whole spectral range in 7 out of 9 SS: H1-36, showing a flatter continuum whose shape is due to a cooler (700-800K instead of 1000K) dust (Allen 1983; Paper I) and BI, where no traces of amorphous silicate bands are recognizable and the spectrum appears almost dust featureless; nonetheless, it is worth noticing that for this last object the poor S/N of the ISO spectrum doesn't allow to conclusively state about the actual presence of some dust features.
\subsubsection{The $\sim$10 \mum \, silicate profile}
\begin{figure}[!hp]
\centering
\includegraphics[width=0.45\textwidth]{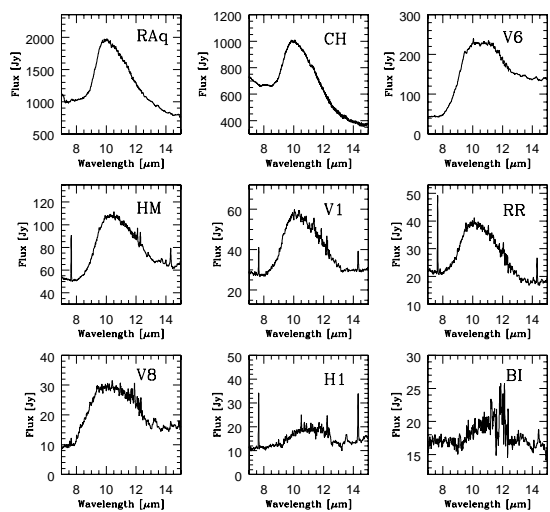}
\caption{The 7-15 \mum \, range. The amorphous silicate profile at $\sim$10 \mum \, is dominant in 7 out of 9 SS.\label{fig:sil10}}
\end{figure}

\begin{figure}[hp!]
\centering
\includegraphics[width=0.45\textwidth,]{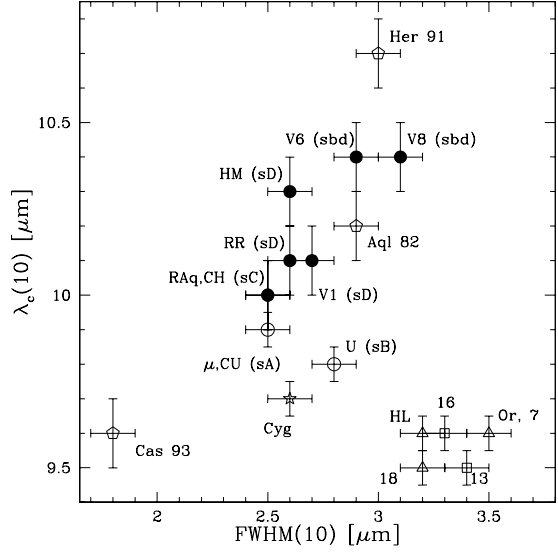}
\caption{Effect of environment on the 10 \mum \, silicate profile. Filled circles: SS; open pentagon: Nov\ae{} (Aql82, Her 91, Cas 93 - Evans et al. 1997); open circles: circumstellar (CU Cep, U Aur, $\mu$ Cephei; Russell et al. 1975); open star: diffuse medium (Cyg OB2 n. 12); open triangles: YSO and TMC (HL Tau, Taurus-Elias 7 and 18) and Orion Trapezium (Gillett et al. 1975); open squares: quiescent TMC (Taurus-Elias 16 and 13).\label{fig:cor1}}
\end{figure}

The SS displaying the 10 \mum \, band might be related, on the basis of the profile shape, to the S00 classification scheme for AGB stars. A direct comparison of our Fig. \ref{fig:sil10} with Fig. 1 of S00 shows that R Aqr and CH Cyg have a profile typical of the \textit{AGB silicate C} group, HM Sge, V1016 Cyg and RR Tel belong to the \textit{AGB silicate D} class, and V627 Cas and V835 Cen lie in the \textit{AGB broad feature} group. No evidence for the $\sim$13 \mum \, feature sometime present in AGB stars is found. In Tab. \ref{tab:dust} and Fig. \ref{fig:cor1} we investigate the effect of SS environments on the band profiles, plotting the FWHM vs. the central wavelength of the peak $\lambda_c$. SS appear in a specific locus of the diagram between the proper circumstellar objects and interstellar environments, being this a clue to an increasing reprocessing of grains across the systems (see below).

It is worth noticing that the only two stars that exhibit the \textit{silicate C} group profile (R Aqr and Ch Cyg) and that lie, in Fig. \ref{fig:cor1}, nearest to the proper circumstellar environments (open circles) display just one dust shell (T$_d$ $\sim$ 1000K) and share similar characteristics both in the continuum SED (Paper I) and in the physical conditions of the ionized nebul\ae{} (\S 3.2.2). This might be just a coincidence: therefore we would like to further investigate the potential implications of such a link in the symbiotic system in a forthcoming, dedicated paper (Contini 2006, in preparation). \\
The three stars (HM Sge, V1016 Cyg, and RR Tel) known to have a very complex network of wakes and shock fronts and many emission lines on top of the continuum, as evidence of strong contributions from the nebul\ae{}, display a $\sim$10 \mum \, profile belonging to the same group, the \textit{AGB silicate D}. Even if the FWHM of the 10 \mum \, band appears practically identical to that of typical stellar objects, the central wavelength of the peak is clearly at longer wavelengths ($\lambda_c$ $\geq$ 10.1 \mum), probably suggesting irregularity in shapes of dust grains (Min et al. 2003). These profiles are well fitted by a combination of Mg-rich silicates and porous amorphous alumina and are interpreted in the light of a dust condensation sequence that can still be related to the specific condition of the isolated AGB star environments (Tielens 1990, Sloan \& Price 1998, Speck et al. 2000). \\
As previously mentioned, V627 Cas and V835 Cen belong to the \textit{AGB broad feature} group. In this case not only the $\lambda_c$ appears at longer wavelength with respect to the other classes, but also the FWHM is larger (FWHM $\sim$ 2.9-3.1 \mum). One of the interpretation of the larger width of the \textit{AGB broad feature group} profile suggests a $\sim$2 \mum-wide excess centred at 11.5 \mum \, which has been attributed to the shifted peak of larger micrometer-sized grains (Jaeger et al. 1994). The presence of these larger grains could be a clue to crystalline silicates (van Diedenhoven et al. 2004), which have been observed in the D'-type SS HD330036 (Paper II). New high resolution and high S/N observations would be welcome considering the other species of silicates disclosed by the ISO "revolution". As a matter of fact these data might reveal the sharp substructures due to crystalline silicates which appear superimposed on the amorphous silicate broad features. A detailed analysis of the crystalline band would then be helpful in determining the physical parameters of these environments, thanks to derived estimates of mass-loss rates (Kemper et al. 2001), 
dust temperature (Bowey et al. 2002) and dust geometrical distribution (Molster et al. 2002a,b,c).
On the other side, new laboratory measurements are necessary for a further verification of this insight under varying environmental parameters, allowing for a better correlation between the profile shape variations and the astrophysical condition parameters.

To summarize,  from this spectral comparison it follows on that dust in D-type SS appears, at least in some cases, different from the dusty environment of ''normal'' isolated late-type evolved stars. In particular, the sample seems to indicate a kind of sequence that moves towards objects with multiple and complicated structures of nebul\ae{} and shells. We suggest that for these systems (the ones that differ more from typical circumstellar environment on the basis of Fig. \ref{fig:cor1}) the different profile features can be interpreted in the light of mixed properties of the grains, being the grains responsible of the bulk of the emission both in the inner "circumstellar shell" and in the outer, circumbinary one (Paper I).
The symbiotic behavior seems thus to play a dominant role not only in the dynamical parameters such as location and distribution of dust by affecting both the shell radii and shapes, but also in the chemical properties of the grains, which experience physical conditions both of circumstellar and ISM environments. The signature of dust chemical disturbance due to symbiotic activity should then be assigned to the outer shells of the whole system, where the environmental conditions for grain processing and even crystallization can, in some cases, be achieved (Paper II).
\subsubsection{The $\sim$18 \mum \, silicate profile}
\begin{figure}[hp!]
\centering
\includegraphics[width=0.45\textwidth,]{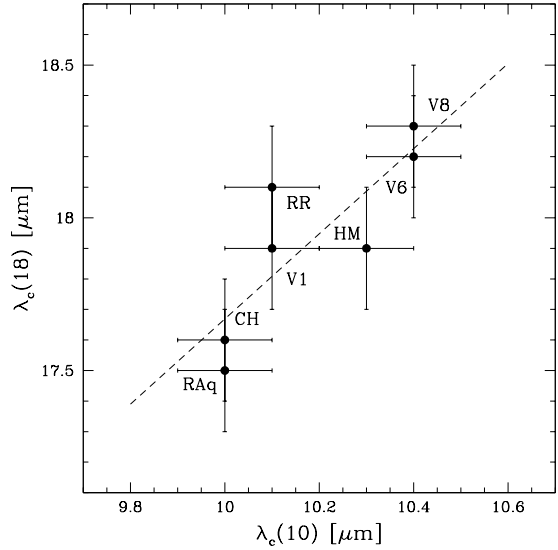}
\caption{The peak wavelength of the 10 \mum \, vs. 18 \mum \, amorphous silicate bands. The dashed lines is the best linear fit with the data.\label{fig:cor2}}
\end{figure}

Fig. \ref{fig:sil} shows that several SS have a second bump at $\sim$ 18 \mum, usually attributed to Si-O bending mode in amorphous silicates. Even in this case, many emission lines are seen on top of the continuum and of the dust band.\\ The strength of the $\sim$18 \mum \, profile decreases along with that of the $\sim$10 \mum\ one; as these two features arise from different vibrational modes of the same chemical bound (Si-O), we wonder if the dust bands answer in a somewhat correlated ways to the same environmental conditions, probably periodically varying in reaction to the Mira periodic pulsations. \\
To verify this idea we plot in Fig. \ref{fig:cor2} (see also Tab. \ref{tab:dust}) the 10 \mum \, vs. 18 \mum \, $\lambda_c$. Interestingly, we find a clear increasing correlation between these two observational parameters: the longer the $\lambda_c$(10), the longer the $\lambda_c$(18). The least square method was performed in order to find the best linear fit to the data: 
$\lambda_c(18)=1.39 \lambda_c(10)+3.73$.

Unfortunately the sample is poor in number of objects, and limited to D-type SS. We might expect that a similar behavior of amorphous silicate dust bands could also be present both in other circumstellar envelopes and galactic environments. It would be interesting if, 
after verified and constrained on a larger sample of different astrophysical environments, 
this relationship will gain some chemical and physical significances on the basis of further laboratory investigations.
\subsection{Dust and gas conditions in the ionized nebul\ae{}}
In SS two main shocks develop from collision of the winds from the stars,
one propagating outwards the system, and the other propagating in reverse
towards the WD, between the stars. The nebulae downstream of the shock fronts are heated and
ionized by the photoionizing flux from the hot star and by shocks.
In order to include this theoretical framework in our model, we complete our investigation on dust features in SS environments by calculating dust and gas conditions downstream of the shocked
nebulae. We will constrain the gas and dust physical conditions
by modeling the strongest emission line ratios.
The IR spectra show lines from a large range of
ionization levels, including [FeVII] and [SiVII].
Radiation from the WD cannot lead to gas temperatures $\geq$ 2-3 10$^4$ K
in the nebulae, while the collision processes heat the gas to temperatures
which depend on the shock velocity (T$\propto$ V$_S^2$). The observed lines
are forbidden, therefore hardly emitted from the shells where very high electron
densities lead to collisional deexcitation.
Moreover, the high ionization level lines indicate high shock velocities
characteristic of the nebula downstream of the reverse shock.

The models which best fit the line ratios are calculated by SUMA (Viegas \& Contini 1994; Contini 1997), a code that simulates the physical conditions of an emitting gaseous cloud under the coupled effect of ionization from an external radiation source and shocks, and in which line and continuum emission from gas are calculated consistently with dust reprocessed radiation as well as with grain heating and sputtering processes. Its calculation models have been already applied to several symbiotic stars, e.g. AG Peg (Contini 1997, 2003), HM Sge (Formiggini, Contini \& Leibowitz 1995), RR Tel (Contini \& Formiggini 1999), He2-104 (Contini \& Formiggini 2001), R Aqr (Contini \& Formiggini 2003), as well as to nova stars (V1974, Contini et al. 1997 - T Pyx, Contini \& Prialnik 1997) and supernova remnants (e.g. Kepler's SNR, Contini 2004).
The composite models are characterized by the parameters referring to
the radiation flux (the temperature of the hot star, \Ts, and the ionization
parameter, U), by those representing the shock (the shock velocity,
\Vs, the preshock density, \n0, and the magnetic field, \B0), and by those related with dust (the dust-to-gas ratio, $d/g$  and the initial grain radius, \agr). The relative abundances to H of He, C, N, O, Ne, Mg, Si, S, Ar and Fe are also accounted for.
\subsubsection{HM Sge: a test case}
We refer to the emission line measurements of SWS spectra for HM Sge by Schild et al (2001). In Table \ref{tab:hmlines} the observed line ratios to [NeVI]
are compared with model results.
We consider first the Ne lines, [NeVI]7.65, [NeV]14.32, [NeIII]15.55, and 
[NeII]12.82,  representing four different ionization levels of the same element.
[NeVI] is the strongest line while [NeII] is very weak, indicating that
the temperature of the gas is high enough across  the whole nebula 
to prevent   recombination.
This constrains the model to matter-bound,
suggesting that the emitting nebula is located between the stars, downstream of the  
shock which propagates in reverse towards the hot component. 
The relatively high velocities indicated by the FWHM of the lines ($\sim$ 500
\kms,  Table \ref{tab:hmlines}) confirm that we are dealing with the reverse shock.
Such high velocities are in agreement with a plasma temperature
of 10$^{6.6}$ K deduced from the X-ray observations by M\"urset et al. (1997).
The profile of the density and temperature downstream which result from the
hydrodynamical equations
determine the stratification of the ions and thus the  intensity ratios
of the lines.
The best fit to the data is obtained with a preshock density \n0=5.10$^5$ \cm3 which leads to densities downstream n $\sim$ 10$^8$
\cm3, and with \Ts=160 000 K and U=1. The corresponding ionizing photon flux
is F$_{\nu}$= 6 10$^{26}$ cm$^{-2}$ s$^{-1}$. 
If the radius of the ionizing source is R$_{WD}\sim$ 10$^{9}$ cm
and the distance to the nebula is $r$, we have $F_{\nu}$R$_{WD}^2$/$r^2$= U n c,
where c is the velocity of light. We obtain $r$=7.8 10$^{13}$ cm. The geometrical thickness of the nebula, $D$= 1.4 10$^{14}$ cm, which is lower than the binary separation (3 10$^{14}$ cm, Richards et al. 1999) is found by modeling.

\begin{figure}[hp!]
\centering
\includegraphics[width=0.45\textwidth,]{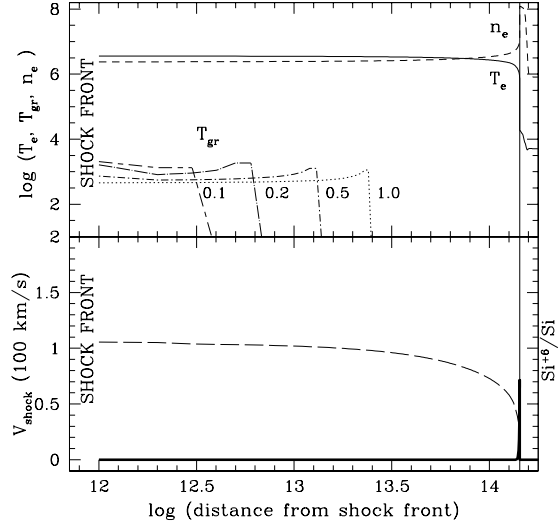}
\caption{Top: the profile of electron density, electron temperature and grain temperature relative to different initial grain size (in \mum) downstream across the reverse shocked nebula in HM Sge. The vertical thin solid line represents the edge of the matter-bound model. Numbers near the grain temperature line indicate the initial grain radius, \agr. Bottom: the profile of shock velocity, \Vs, and the distribution of the Si$^{+6}$/Si fractional abundance (thick solid line).\label{fig:t1}}
\end{figure}

\begin{figure}[hp!]
\centering
\includegraphics[width=0.45\textwidth,]{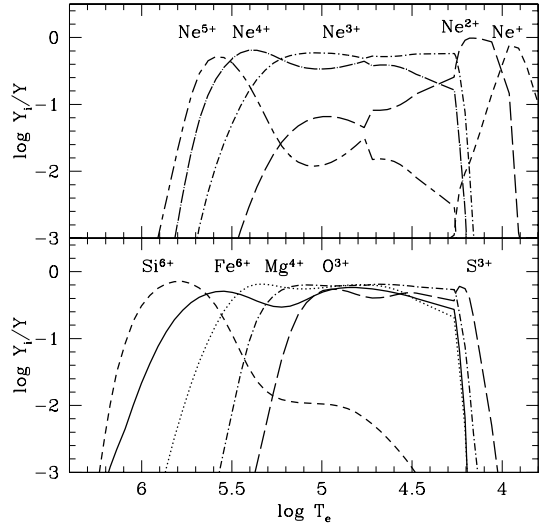}
\caption{The fractional abundance of the ions downstream
as function of the temperature across the nebula for HM Sge.\label{fig:t2}}
\end{figure}

Once the Ne line ratios are reproduced by the model, the relative abundances 
of the elements result from the agreement of the calculated with the observed line
ratios. However, the elements different from Ne are represented by only one line
in the observed spectrum, so the relative abundances are less constrained.\\
In Fig. \ref{fig:t1} top, the profiles of the electron density, \ne, of the electron temperature of the gas, \Te, as well as of the temperatures of dust grains, \Tgr, corresponding to different initial grain radii are shown across the nebula. The vertical thin solid line shows
that most of the recombination region is cut off.
The fractional abundance of the Ne ions  (top)  and of the ions corresponding
to the other observed lines (bottom) are shown in Fig. \ref{fig:t2} as a function
of the electron temperature.\\ 
In Table \ref{tab:abund} our results are compared with those of previous works.
We find O/H and Ne/H  higher than solar by factors of $\sim$ 2.3 and 4.8, respectively, revealing the strong contribution of the WD wind to the nebula between the stars (Nussbaumer \& Vogel 1990, Weidemann 2003). Moreover this allows us to postulate that oxygen had an initial (i.e. pre-1975 outburst) abundance as high as to compensate any depletion caused by subsequent grain formation processes. The other metals such as Mg, Si and Fe, which are easily trapped into grains, are instead depleted relatively to the solar values. Interestingly, the depletion of Fe does not appear correlated with that of Mg and Si, being instead more than twice larger. If iron was incorporated in the silicate lattice it would return to the gas phase along with the magnesium and silicon when the silicate is destroyed (see below); iron then must be trapped in different dust material, maybe in the form of refractory species such as metallic iron or iron oxide, so that it can stay in a solid phase while the magnesium and silicon are returned to the gas phase. This insight, which provides a natural explanation of the unusual depletion pattern observed also in several lines of sight through the interstellar medium (Welty et al. 2001; Sofia et al. 2006) has been already suggested by Sofia et al. (2006) and consistently confirmed by a very recent theoretical study by Min et al. (2006). As a further and complementary proof we deduce our results on the basis of the analysis of atomic emission lines.

The rate of erosion (sputtering) of the grains  can be high for \Vs $\geq$ 200 \kms,
leading to a complete destruction of the grains. Notice that even the largest ones (\agr= 1 \mum) are sputtered within a relative small distance from the shock front (Fig. \ref{fig:t1}, top). Mg, Fe and particularly Si should therefore return to the gaseous phase before the temperature drop to the values suitable to the line emission. This indicates that other effects must be considered, e.g. the lifetime of grain survival which depends on the distance downstream covered by the grains (Shull 1978, Dwek 1981). Small grains are stopped after distances of $\sim$ 7 10$^{12}$ cm, while large grains after $\sim$ 3 10$^{13}$ cm (Contini et al. 2004) which are of about the same order of the sputtering distances shown in Fig. \ref{fig:t1}.

An intriguing issue related to grain processing is that the return of silicon to the gas phase might take place in the molecular form (SiO, SiO$_{2}$), hence explaining the apparent contradiction between the high efficiency of sputtering processes we deduce from the shocks and the observed depletion calculated by the atomic lines. \\
It is known that SiO is a very reliable tracer of shocked gas in many astrophysical environment, and its enhancement in the gas phase has been already observed both in the bipolar outflows of young stellar objects (Martin-Pintado et al. 1992; Schilke et al. 1997) and in galaxies (Garcia-Burillo et al. 2000, 2001b; Usero et al. 2006). Moreover, there is growing evidence that different molecul\ae{} trace distinctly different velocity regimes in shocks, with SiO associated with more energetic events, i.e., those potentially more efficient in processing dust grains (Garay et al. 2000). Although qualitatively similar, shocks characterized by different velocity regimes are expected to develop to a different extent dust grains in molecular gas: fast shocks can destroy the grain cores, liberating refractory elements to the gas phase, while slow shocks could only be able to process the icy grain mantles. One can then conclude that an increase in the typical velocity regime of shocks will favor an enhancement of the abundance of SiO in the shocked gas.\\
The shock velocity characteristic of these processes, i.e. the threshold value above which the dissociation of SiO takes place, are usually \Vs$ \leq$ 60-70 \kms (Martin-Pintado et al. 1992). At the bottom of Fig. \ref{fig:t1} the profile of the shock velocity downstream which results from the solution of the continuity equations for our model is shown, along with the [SiVII] line formation region. As one can see, at distances sufficiently large from the shock front the velocity drops to \Vs $<$ 100 \kms, indicating suitable conditions for molecular shock chemistry.\\
The interpretative framework is still in progress under many aspect, and it needs observational data to be more constrained. Unfortunately, so far, there is a complete lack of spectral observations for SS in the $mm$ spectral region. We then call for $mm$ observations at SiO transition frequencies, e.g. $\sim$ 217 GHz J=5-4 ($\lambda$=1.38 $mm$), $\sim$ 130 GHz J=3-2 ($\lambda$=2.3 $mm$), $\sim$ 86 GHz J=2-1 ($\lambda$=3.4 $mm$), in order to trace the shock spatial profile within and around the symbiotic nebul\ae{} and to reveal up to what degree the sputtering processes are at work. 

Consider now the nebula downstream of the shock front expanding outwards the system which 
is characterized by a higher $D$, lower velocities and lower densities, hence being the source of the rich optical and UV spectra reported and studied extensively by previous works (e.g. Nussbaumer \& Walder 1993, Formiggini et al. 1995). The grains can survive to destructive processes and contribute to the IR emission even at larger wavelengths, because of the lower temperature (\Tgr $\sim$ 300K, Paper I) of dust. \\
The overall conditions of dust and grains within the outer expanding nebul\ae{} are approaching those of the ISM up to a complete merge, indicating that the signature of dust chemical disturbance due to symbiotic activity, i.e. the chemical and physical reprocessing of the grains suggested on the basis of the spectral analysis, might take place and should be looked for in these circumbinary structures.
\subsubsection{The other objects}
Fig. \ref{fig:sil} shows that some objects of the sample have very similar spectra, both in the dust and in the emission lines features, while some others lack highly ionized lines (e.g. R Aqr, CH Cyg) and/or dust band (e.g. BI Cru, H1-36).

For objects with an evident emission line spectrum the Ne spectral line ratios, which are independent from relative abundance effects, are compared in Table \ref{tab:lines}. It can be noticed that the general trend is very similar to that found for HM Sge, except RR Tel which shows [NeIII]/[NeVI] lower by a factor $\leq$ 2, and H1-36, which displays a very bright [OIV]25.89\mum\ emission line, as well as no evident dust spectral feature (see \S 3.1). In Table \ref{tab:fwhm} we present the FWHM measured for the brightest lines. The derived velocities indicate that also for these objects, as for HM Sge, we are dealing with the inner shocked nebula, which emits the detected atomic emission lines and where dust grains can hardly survive the harsh environmental conditions. Considerations similar to those reported in the previous section (\S 3.2.1) about the molecular emission are still applicable.

It is noteworthy that R Aqr and CH Cyg, known to have just one hot (\Tgr $\sim$ 1000K) and circumbinary dust shell, do not show emission lines from highly ionized elements (Paper I). A plausible explanation for the absence of such strong atomic features could lie in the high density of the inner shocked nebula, up to 10$^9$ \cm3 (Contini 2006, in preparation), that prevents the formation of forbidden emission lines.

Unfortunately for some objects ISO spectra are too noisy to reveal fainter spectral feature (V835 Cen, BI Cru). Furthermore, there are still insufficient data in some spectral ranges for a consistent and thorough analysis, and even for concluding about the actual symbiotic nature of some stars, as V627 Cas (Belczy{\'n}ski et al. 2000, Angeloni et al. 2006, in preparation). Thus we postpone to forthcoming papers a detailed discussion of the single objects.

\section{Concluding remarks}
We have investigated the IR spectral features of a sample of D-type symbiotic stars in order to constrain the emitting properties of coupled dust-gas particles across the whole systems. Having extracted the basic observational parameters of both dust bands and atomic emission lines from unexploited ISO data and having compared them with those from other astronomical sources, we have attempted to highlight the effect of the environment on grain chemistry and physics, in particular whether dust grains can survive within the shocked nebul\ae{} and how their chemical properties might be affected by several depletion and sputtering patterns.\\
The main conclusions of this work are the following:
\begin{itemize}
\item The strong amorphous silicate emission features at $\sim$10 \mum\ and $\sim$18 \mum \, are found in 7 out of 9 objects of the sample. A qualitative analysis of the $\sim$10 \mum\ dust band indicates that the profile shape is similar to that of "standard" isolated AGB stars. Nevertheless a quantitative comparison of $\lambda_c$ vs. FWHM for the 10 \mum\ silicate profile with other astronomical sources shows that SS appear between the characteristic circumstellar objects and interstellar environments, which is believed to be a clear clue of an increasing, ongoing reprocessing of grains. 
\item A correlation between the central wavelength of the $\sim$10 \mum\ and $\sim$18 \mum \, dust bands is found. It is likely that the dust bands answer in a somewhat correlated ways the same environmental conditions, probably periodically varying in response to the Mira periodic pulsations if the dust grains responsible of the bulk of the emission are mainly located in the hot, circumstellar shell. Both new high resolution and high S/N observations, as well as new laboratory measurements under varying environmental parameters, are necessary for a further verification of this insight, even thanks to larger samples.
\item The modeling of IR spectral lines allowed to constrain the emitting properties of coupled dust-gas particles within the shocked nebul\ae{}, showing that sputtering processes are efficiently at work in the inner nebul\ae{} where high shock velocities are reached. The signature of dust chemical disturbance due to symbiotic activity should then be looked for into the outer, expanding, circumbinary shells where the environmental conditions for grain processing and even crystallization might in some cases be achieved.
\item Both the unusual depletion values and the high sputtering efficiency could be explained by the overabundant presence of SiO molecul\ae{}, known to be a very reliable shock tracer, hence making SS interesting targets for $mm$ observations.
\end{itemize}
\section*{Acknowledgments}
We are grateful to the referee, Nye Evans, for many valuable comments. We would like to thank A. Cicakova for reading the manuscript. RA is thankful to the School of Physics \& Astronomy of the Tel Aviv University for their kind hospitality.

\clearpage

\begin{table*}
\centering \caption{$\lambda_c$ and FWHM of the 10 \mum\ silicate profile, $\lambda_c$ of the 18 \mum\ silicate profile, and relative uncertainties [\mum] for the objects in Figs. \ref{fig:cor1} and \ref{fig:cor2}. \label{tab:dust}}
\begin{footnotesize}
\begin{tabular}{c c c c c c c}
\hline
\hline
Object & $\lambda_c$(10) & err $\lambda_c$(10) & FWHM(10) & err FWHM(10) & $\lambda_c$(18) & err $\lambda_c$(18)\\
\hline
\textbf{Symbiotic}&&&&&\\
R Aqr & 10.0 & $\pm$0.1 & 2.5 & $\pm$0.1 & 17.5 & $\pm$0.2\\
CH Cyg & 10.0 & $\pm$0.1 & 2.5 & $\pm$0.1 & 17.6 & $\pm$0.2\\
V627 Cas & 10.4 & $\pm$0.1 & 2.9 & $\pm$0.1 & 18.2 & $\pm$0.2\\
HM Sge & 10.3 & $\pm$0.1 & 2.6 & $\pm$0.1 & 17.9 & $\pm$0.2\\
V1016 Cyg & 10.1 & $\pm$0.1 & 2.7 & $\pm$0.1 & 17.9 & $\pm$0.2\\
RR Tel & 10.1 & $\pm$0.1 & 2.6 & $\pm$0.1 & 18.1 & $\pm$0.2\\
V835 Cen & 10.4 & $\pm$0.1 & 3.1 & $\pm$0.1 & 18.3 & $\pm$0.2\\
\textbf{Nov\ae{}}&&&&&\\
Aql 82 & 10.2 & $\pm$0.1 & 2.9 & $\pm$0.1 & - & -\\
Her 91 & 10.7 & $\pm$0.1 & 3.0 & $\pm$0.1 & - & -\\
Cas 93 & 9.6 & $\pm$0.1 & 1.8 & $\pm$0.1 & - & -\\
\textbf{Circumstellar}&&&&&\\
$\mu$ Ceph & 9.9 & $\pm$0.05 & 2.5 & $\pm$0.1 & - & -\\
CU Cep & 9.9 & $\pm$0.05 & 2.5 & $\pm$0.1 & - & -\\
U Aur & 9.8 & $\pm$0.05 & 2.8 & $\pm$0.1 & - & -\\
\textbf{Diffuse medium}&&&&&\\
Cyg OB2 n.12 & 9.7 & $\pm$0.05 & 2.6 & $\pm$0.1 & - & -\\
\textbf{YSO, Orion Trap., TMC}&&&&&\\
HL Tau & 9.6 & $\pm$0.05 & 3.2 & $\pm$0.1 & - & -\\
Orion Trap. & 9.6 & $\pm$0.05 & 3.5 & $\pm$0.1 & - & -\\
Taurus-Elias 7 & 9.6 & $\pm$0.05 & 3.5 & $\pm$0.1 & - & -\\
Taurus-Elias 13 & 9.5 & $\pm$0.05 & 3.4 & $\pm$0.1 & - & -\\
Taurus-Elias 16 & 9.6 & $\pm$0.05 & 3.3 & $\pm$0.1 & - & -\\
Taurus-Elias 18 & 9.5 & $\pm$0.05 & 3.2 & $\pm$0.1 & - & -\\
\hline
\end{tabular}
\end{footnotesize}
\end{table*}
\begin{table*}
\centering \caption{emission lines, normalized to [NeVI] 7.65 \mum, in the SWS spectrum of HM Sge.\label{tab:hmlines}}
\begin{tabular}{c c c c c c}
\hline
\hline
$\lambda$ [\mum] & ID & iP [eV] & FWHM [\mum] & observed$^a$ & model$^b$\\
\hline
4.53 & [ArVI] & 91.01 & 0.01 & 0.09 & 0.09\\
5.61 & [MgV] & 141.27 & 0.04 & 0.19 & 0.18\\
6.49 & [SiVII] & 246.52 & 0.03 & $<$0.07 & 0.07\\
7.65 & [NeVI]$^c$ & 157.93 & 0.03 & 1 & 1\\
9.53 & [FeVII] & 124.98 & 0.04 & 0.05 & 0.05\\
10.51 & [SIV] & 47.22 & 0.04 & 0.04 & 0.04\\
12.81 & [NeII] & 40.96 & 0.07 & $<$0.07 & 0.001\\
14.32 & [NeV] & 126.21 & 0.06 & 0.13 & 0.12\\
15.55 & [NeIII] & 63.45 & 0.06 & 0.08 & 0.08\\
24.31 & [NeV] & 126.21 & 0.11 & 0.06 & 0.04\\
25.89 & [OIV] & 77.41 & 0.10 & 0.09 & 0.08\\
\hline
\end{tabular}
\flushleft
\scriptsize
a: adapted from Schild et al. 2001;\\
b: input parameters: \Vs=500\kms; \n0=5 10$^5$ \cm3; \B0=10$^{-3}$ G; d/g=4 10 $^{-5}$ (by mass); \agr=1, 0.5, 0.2 \mum; D=1.74 10$^{14}$ cm; \Tgas=1.85 10$^4$ K; \Ts=1.6 10$^5$ K; U=1;  \\
c: absolute flux: 74 $\pm$ 0.6 10$^{-12}$ erg cm$^{-2}$ s$^{-1}$.
\end{table*}
\begin{table*}
\centering \caption{the comparison of relative abundances for HM Sge.\label{tab:abund}}
\begin{tabular}{c c c c c c}\\
\hline
\hline
element & model & NV90 & solar\\
\hline
He/H & 0.11 & 0.10 & 0.085\\
O/H & 1.5(-3) & 1.0(-3) & 6.6(-4)\\
Ne/H & 4.0(-4) & 4.0(-4) & 8.3(-5)\\ 
Mg/H & 1.5(-5) & 1.0(-4) & 2.6(-5)\\
Si/H & 8.0(-6) & 2.0(-4) & 3.3(-5) \\
S/H & 3.6(-5) & 1.0(-4) & 1.6(-5)\\
Ar/H & 6.3(-6) & 6.0(-6) & 6.3(-6)\\
Fe/H & 6.0(-6) & 3.0(-5)  & 4.0(-5)\\
\hline
\end{tabular}
\flushleft
\scriptsize
NV90: Nussbaumer \& Vogel (1990).
\end{table*}
\begin{table*}
\centering \caption{the brightest Ne emission lines$^a$ visible on top of continuum.\label{tab:lines}}
\begin{tabular}{c c c c c c}\\
\hline
\hline
Symb & [NeV]$_{14.32}$& [NeIII]$_{15.55}$ & [NeV]$_{24.31}$ \\
\hline
R Aqr  &  -&-&-& \\
CH Cyg &  -&-&-& \\
V627 Cas  &-&-&\\ 
HM Sge$^b$  & 0.13 & 0.08 & 0.06\\ 
V1016 Cyg & $<$0.3 & $<$0.1 &$<$0.2\\
RR Tel  & $<$0.2 & $<$0.05 & $<$0.1 \\
BI Cru &-&-& \\
V835 Cen  & $<$0.05 &-&\\
H1-36 & $<$0.5 & $<$0.2 & $<$0.15 \\
\hline
\end{tabular}
\flushleft
\scriptsize
a: normalized to [NeVI] emission line  @ 7.65 \mum; \\
b: adapted from Schild et al. 2001.
\end{table*}
\begin{table*}
\centering \caption{FWHM [\mum] of the brightest atomic emission lines.\label{tab:fwhm}}
\begin{tabular}{c c c c c c}\\
\hline
\hline
Symb & [NeVI]$_{7.65}$  & [NeV]$_{14.32}$& [NeIII]$_{15.55}$ & [NeV]$_{24.31}$ &[OIV]$_{25.89}$\\
\hline
R Aqr  &  -&-&-&-&- \\
CH Cyg &  -&-&-&-&- \\
V627 Cas  & -&-&-&-&-\\
HM Sge & 0.03 & 0.06 & 0.06 & 0.11 & 0.10\\
V1016 Cyg  &   0.03 & 0.05& 0.05 & 0.11& 0.11\\
RR Tel  & 0.03 & 0.06& 0.06 & 0.09& 0.10 \\
BI Cru &-&-&-&-&- \\
V835 Cen & 0.03 & 0.06 &-&-&-\\
H1-36 & 0.03 & 0.05 & 0.06 & 0.10& 0.11 \\
\hline
\end{tabular}
\end{table*}

\clearpage


\end{document}